\documentclass[10pt]{article}
\usepackage{graphicx}
\usepackage{amsmath}
\makeatletter

\begin{document}
\title{\textbf{ Phenomenological approach for describing environment dependent
growths}}
\author{\textbf{Dibyendu Biswas$^1$}\footnote{dbbesu@gmail.com} \textbf{and Swarup Poria$^2$}\footnote{swarupporia@gmail.com}\\
\small{$^1$Department of Basic Science, Humanities and Social Science (Physics)}\\\small{ Calcutta Institute of Engineering and Management }\\
\small {24/1A Chandi Ghosh Road, Kolkata-700040, India}\\
\small{$^2$Department of Applied Mathematics, University of Calcutta}\\
\small {92 Acharya Prafulla Chandra Road, Kolkata-700009, India}}

 \date{}
\maketitle
\begin{center}
{\bf{Abstract}}\\
\end{center}
Different classes of phenomenological universalities of environment dependent growths have been proposed. The logistic as well as environment dependent West-type allometry based biological growth can be explained in this proposed framework of phenomenological description. It is shown that logistic  and environment dependent West-type growths are phenomenologically identical in nature. However there is a difference between them in terms of coefficients involved in the phenomenological descriptions. It is also established that environment independent and enviornment dependent biological growth processes lead to the same West-type biological growth equation. Involuted Gompertz function, used to describe biological growth processes undergoing atrophy or a demographic and economic system undergoing involution or regression, can be addressed in this proposed environment dependent description. In addition, some other phenomenological descriptions have been examined in this proposed framework and graphical representations of variation of different parameters involved in the description are executed.


\section{Introduction}
A classification scheme in universality classes of broader categories of phenomenologies, belonging to different disciplines, is emerging as a useful tool for the purpose of recognition of a characteristic feature of a system and cross-fertilization among different branches of science. Over last few decades, considerable amount of effort has been devoted to describe temporal evolution of systems occurring in physics, biology, statistics and economics. Understanding of non-linear, complex systems in the field of ecology, biology, physics, economics and other branches of social sciences requires macroscopic and microscopic investigation of the system. Prediction of the behaviour of a non-linear system based on microscopic investigation is often very difficult because of the fact that it includes a lot of parameters in the calculation. In case of macroscopic approach, the significant changes may lead to some new features. But as they take place at the microscopic level, they are often ignored. A mid-way description of the system, termed as mesoscopic approach, may be helpful in some cases {\cite {Delsanto2000,Delsanto2003,Delsanto1998}}, but the coordination among different level of description may not be easy {\cite {Delsanto2005}}. At this conjuncture, phenomenological approach to the system is emerging as a useful interdisciplinary tool to describe the temporal evolution of systems.\\
The proposal of different types of universality classes prepares the background of studying a dynamical system based on the phenomenological approach {\cite {Delsanto2005,Guiot,Delsanto2007,West1997,Gillooly,West2001,West2004}} . In each class, a characteristic parameter of different systems shares common set of  properties. The field of application and the nature of the variables involved are completely ignored in this approach. This new approach is used for the analysis of experimental datasets and cross-fertilization among different fields; e.g. physics, engineering, medical science and different branches of social sciences. In fact, this method indicates similarities to be detected among datasets in totally different fields of application and serves as a magnifying glass upon them, enabling one to extract all the available information in a generalized way. \\
The main aim of the phenomenological approach {\cite {Castorina}} is the formulation of the family of classes (UN), which at the level N = 0 corresponds to autocatalytic processes. At the level N = 1 (U1), it produces the Gompertz law {\cite {Gompertz}} which was derived first empirically. It has been used to describe most of the diversified growth phenomena, for more than a century. The class U2 includes the model of West and his collaborators
{\cite {West2001,West2004}}. Different types of applications of the classes U1 and U2 have been reported in diversified and unrelated fields ; eg. auxology {\cite {Delsanto2008}}, non-linear elasticity {\cite {Delsanto2009}}, oncology {\cite {Gliozzi, Gliozzi2010}} and fracture dynamics {\cite {Pugno}}. It is also found useful to detect the phase of inflation growth {\cite {Iordache}}. In nonlinear problems, this approach shows the nonscaling invariance to be extracted by means of suitable redefined fractal-dimensioned variables {\cite {Delsanto2008}}. An extension of this phenomenological approach in terms of complex variable(s), termed as complex univeralities of growth, is used to explain concurrent growth of phenotypic features of a system {\cite {Barberis2010, Delsanto2011}}.\\
It is found that different systems proposed by Castorina $et$ $al.$ are capable of attaining carrying capacity of the corresponding system. The carrying capacity here is controlled by phenomenological coefficients indicating growth mechanisms of the system. Therefore, in these cases carrying capacity does not depend on environmental conditions. The carrying capacity and the maximum attainable value of the dynamical variable $y$ are synonymous in these cases. But there are different growing systems which are unable to reach carrying capacity {\cite{valdar2006}}. At the same time, in Valdar's words {\cite{valdar2006}}, \textquotedblleft growth rate is an adaptive mechanism that responds to the environmental conditions, which enter as the initial conditions \textquotedblright. Castorina $et$ $al.$ {\cite {Castorina}}  have not considered these essential aspects of growth process in their proposed U1 and U2 classes. \\
One of the important factors in the classification scheme of the phenomenological universalities is the initial conditions. Therefore, new initial conditions may be considered to define a new class whose growth is effected by environmental conditions. In the present communication, following the formalism developed by Castorina et. al.{\cite {Castorina}}, a new type of phenomenological description, based on proposed initial conditions, is considered to describe the temporal evolution of the system. In this type of description of the system, environment dependent West-type biological growth and logistic type growth pattern can be addressed in the same framework. That there is no difference in the underlying phenomenological descriptions in between environment dependent West-type and logistic-type growth pattern is established. The only difference is in terms of the value of coefficients involved in phenomenological description. A switching between environment dependent West-type and logistic-type growth pattern is shown in a proper way. It is also shown that the biological growth following West-type growth equation proposed by Castorina $et$ $al.$ {\cite {Castorina}} is different from the proposed biological growth governed by environment dependent West-type growth equation. Involuted Gompertz function, used to analyze growth of a biological system undergoing atrophy or a demographic or economic system following involution or regression {\cite {Molski}}, can also be addressed from the proposed phenomenological point of view. In addition, some phenomenological descriptions in this framework have been explained with its minute features, along with graphical representation.\\
The paper is organized as follows: In Sec. II, we would propose a phenomenological description of environment dependent growth processes. In this connection, the classification scheme of phenomenological universalities proposed by Castorinal $et$ $al.$ {\cite {Castorina}} is discussed in brief. Different aspects of proposed phenomenological classes would be considered in Sec. III. Finally we would conclude with our results in Sec IV.
\section{The phenomenological description}
In a generalized way, any growth phenomenon may be described by a simple relation as given below,
\begin{equation}
\frac{dY(t)}{dt}=\alpha(t)Y(t)
\end{equation}
where $\alpha(t)$ represents specific growth rate of a given dynamical variable $Y(t)$. $Y(t)$ may vary with some other independent characteristic variable of the system. But, the evolution of the system with respect to time is considered in the phenomenological approach, treating others as constant. Therefore, an ordinary differential equation, instead of a partial one, will serve the purpose. Now, the equation (1) can be expressed in terms of nondimensional variables, by introducing three nondimensional variables $\tau=\alpha(0)t$, $y(\tau)=Y(t)/Y(0)$ and $a(\tau)=\alpha(t)/\alpha(0)$, as
\begin{equation}
\frac{dy(\tau)}{d\tau}=a(\tau)y(\tau)
\end{equation}
Now, the time variation of $a(\tau)$ is defined through the function,
\begin{equation}
\varphi(a)=-\frac{da}{d\tau}
\end{equation}
The explicit form of $\varphi(a)$ will generate a variety of growth patterns. $\varphi (a)$ can be expressed in terms of a power series as,
\begin{equation}
\varphi(a)={\sum_0^\infty} b_n a^n
\end{equation}
From equation (2) and (3), the following relation can be established,
\begin{equation}
\ln y=-\int \frac{ada}{\varphi}+ constant
\end{equation}
Equation (5) can also be expressed in the following form,
\begin{equation}
\ln y=\int ad\tau+ constant
\end{equation}
The growth processes may be endogenous or exogenous by nature. Both of them can be addressed with the help of this classification scheme of the phenomenological universalities of growth {\cite{Delsanto2011}}. Equation (2) and (3) along with equation (4) generate different types of universality classes in the growth processes. One of the important factors in this classification scheme is the initial conditions that may depend upon its environmental conditions. There is no scope to consider such environmental constraints in the initial conditions proposed by Castorina $et.$ $al.$ {\cite{Castorina}}. The initial conditions are independent of the environmental conditions. Different terms of equation (4) represent several growth mechanism of the system {\cite{Castorina}}. But the growth processes are expected to be influenced by the environmental conditions. Here, we propose the initial conditions as $y(0)=1$ and $a(0)=a_0=q(1-\frac{1}{K})$, where $q$  is a constant quantity. $q$ can be treated as the ideal specific growth rate that may not be attained by the system. Ideal specific growth rate may be equal to (one of the) phenomenological coefficient(s). In reality, actual specific growth rate may be less than $q$. It may be due to some constraints imposed by its environment. In this proposed initial conditions, $K$ is another constant representing the interaction of the system with its environment. $K$ may be treated as the carrying capacity that may or may not be attained by the system {\cite{valdar2006}}.
The behaviour of the system corresponding to constant specific growth rate, i.e. $\varphi=0$ ($b_n=0$ for any $n$) is represented by the following differential equation,
\begin{equation}
\frac{dy}{d\tau}=\pm ry
\end{equation}
where, $r$ is a constant quantity.\\
When $r$ is a positive constant quantity, the system shows an exponential growth. The system shows an exponential decay when $\varphi=0$ and $a$ is equal to a constant (less than zero). For the condition $\varphi=b_0$ and $q=b_0$, the system is governed by the following relation,
\begin{equation}
\frac{d^2y}{d\tau^2}=\frac{1}{y}(\frac{dy}{d\tau})^2-b_0y
\end{equation}
With the solution as given below,
\begin{equation}
y=\exp{[-\frac{b_0\tau^2}{2}+b_0(1-\frac{1}{K})\tau]}
\end{equation}
Now, for $b_0=0$, $q=b_1$ and $\varphi(a)=b_1a$ with $n=1$ in equation (4), the growth pattern is governed by the following expression,
\begin{equation}
\frac{dy}{d\tau}=b_1y[(1-\frac{1}{K})-\ln y]
\end{equation}
with the solution,
\begin{equation}
y=\exp {[(1-\frac{1}{K})(1-\exp (-b_1\tau))]}
\end{equation}
The system corresponding to phenomenological class represented by $\varphi=b_0+b_1a$ with all $b_n=0$ for $n>1$; follows,
\begin{equation}
y=exp[(\frac{b_0}{b_1^2}+\frac{a_0}{b_1})(1-exp(-b_1\tau))-\frac{b_0}{b_1}\tau]
\end{equation}
In case of $b_0=b_1=0$ and $\varphi(a)=b_2a^2$ with $n=2$, the corresponding differential equation that shows the growth pattern for the condition $q=b_2$ is expressed as,
\begin{equation}
\frac{dy}{d\tau}=b_2(1-\frac{1}{K})y^{1-b_2}
\end{equation}
with the solution,
\begin{equation}
y=[b_2^2(1-\frac{1}{K})\tau+1]^{\frac{1}{b_2}}
\end{equation}
When $b_0=0$ and $n=2$, $\varphi(a)=b_1a+b_2a^2$, the behaviour of the system for the condition $q=b_1$ is governed by differential equation given as,
\begin{equation}
\frac{dy}{d\tau}=\frac{b_1}{b_2}(y^{1-b_2}+\gamma y^{1-b_2}-y)
\end{equation}
where, $\gamma=\frac{b_2}{K}(K-1)$.\\
The solution of the equation (16) is
\begin{equation}
y=[1+\gamma-\gamma \exp (-b_1\tau)]^{\frac{1}{b_2}}
\end{equation}
This is similar in nature as derived by Castorina et.al. {\cite {Castorina}}, with a little difference in the value of $\gamma$.
\section{Discussions}
In the phenomenological class defined by $\varphi=o$, the system shows an exponential growth or decay based on the value of the constant quantity which is equal to $a$. The system exhibits an exponential decay when the constant quantity is less than zero. One of the systems in which this type of decay is found is radioactive system. The system follows exponential growth when the value of the constant quantity is greater than zero. One of such systems is auto-catalyst driven system. As $a$ is not function of $\tau$ or $y$ in this case, the growth is not endogenous or exogenous by nature. It can be treated as a spontaneous process.\\
For the phenomenological description corresponding to $\varphi=b_0$, the system exhibits a growth at the early stage when $b_0>0$. Thereafter, the system shows an exponential decay and remains asymptotic to the adimensional time axis. The growth rate or decay rate increases with the increase in $K$ when $b_0$ remains constant. The opposite nature is found in the system when $b_0$ is less than zero. The same nature is exhibited by the system with the variation in $b_0$, treating $K$ as a constant. The growth rate or decay rate increases with the increase in $b_0$ when $K$ behaves like a constant. One interesting feature found in this case is that the duration of growth is controlled by the magnitude of $K$, and not by the magnitude of $b_0$. The findings are represented graphically in figure $1$. The maxima of the characteristic growth pattern occurs at $\tau=(1-\frac{1}{K})$. The specific growth rate ($a$) can be expressed explicitly in terms of $\tau$ only. Therefore, the system is exogenous by nature {\cite{Delsanto2011}}.\\
The system corresponding to the phenomenological class represented by $\varphi=b_1a$, shows a growth very similar to Gompertz-law of growth {\cite {Gompertz}}. Figure $2$ shows that the growth rate increases with the increase of $b_1$ when $K$ behaves like a constant quantity. The time required to attain saturation level decreases significantly with the increase of $b_1$ in this case. The growth rate and time required to attain saturation level vary significantly with $K$, even if $b_1$ remains constant. The system behaves like an endogenous system {\cite{Delsanto2011}}. It shows a linear behaviour with respect to time when $b_2=1$ in case of $\varphi=b_2a^2$, as shown in figure $3$. It is also an example of the endogenous system.\\
Equation (12) can be expressed  for the conditions $q=b_0$, $K=b_1$ and $b_0b_1=1$ as,
\begin{equation}
y=exp[\frac{1}{b_1^2}(1-exp(-b_1\tau))-\frac{b_0}{b_1}\tau]
\end{equation}
This is termed as involuted Gompertz function {\cite {Molski}} (shown in figure $4$) generally used to describe the growth process of a biological system undergoing atrophy. Such a situation is found in avian primary lymphoid organs $-$ thymus and bursa of Fabricius as well as in thymus of mammalians {\cite {Molski}}. It is also useful for a demographic and economic system undergoing involution or regression. The growth process corresponding to this class is exogenous {\cite{Delsanto2011}}.\\
Equation (15) can be expressed in the following way,
\begin{equation}
\frac{dy}{d\tau}=C_1y^\sigma - C_2 y
\end{equation}
where, $C_1=\frac{b_1(1+\gamma)}{b_2}$, $\sigma =1-b_2$ and $C_2=\frac{b_1}{b_2}$.\\
This is very similar to the expression proposed by West et. al. for allometric growth of biological systems when $y$ represents mass of the system and $b_2=0.25$ {\cite {West1997,West2004}}. Figure $5$ shows that the time required to attain saturation level decreases with the increase in $b_1$ when $K$ is a constant quantity. The phenomenological class represented by equation (15) shows endogenous growth process.\\
Equation (16) is similar in nature with the expression given by Castorina at. al. {\cite {Castorina}} as
\begin{equation}
y=[1+b-b \exp (-\tau)]^{\frac{1}{b}}
\end{equation}
Therefore, the specific growth rate ($a$) corresponding to equation (15) is represented by,
\begin{equation}
a=\frac{b_1(K-1)}{(1.25K-0.25)exp(b_1\tau) -0.25(K-1))}
\end{equation}
and corresponding mass (when dynamic variable $y$ represents mass) is expressed as,
\begin{equation}
m=[1+\frac{0.25(K-1)}{K}(1-exp(-b_1\tau))]^4
\end{equation}
This is similar in nature with the expression derived by Biswas et.al. {\cite {Biswas}}. It is found that the proposed biological growth leading to the environment dependent West-type growth equation does not attain the carrying capacity. There are different types of growing systems which can not attain carrying capacity {\cite{valdar2006}}.
When $b_2=-1$, equation (15) can be expressed as,
\begin{equation}
\frac{dy}{d\tau}=b_1y(1-\frac{y}{K})
\end{equation}
The solution of the equation (22) is expressed as,
\begin{equation}
y=\frac{K}{1+(K-1)exp(-b_1\tau)}
\end{equation}
This is logistic growth equation that is frequently used to describe different types of growth phenomena in diversified field {\cite {Bose,Gilpin,Sahoo2014,Courtillot}}. Figure $6$  represents the variation of dynamic variable $y$ with time in case of logistic growth for different values of $K$ when $b_1$ and $b_2$ are constant. It is found that the time required to attain saturation level increases slowly with the increase in $K$. The specific growth rate corresponding to logistic growth can be expressed in terms of adimensional time ($\tau$) as,
\begin{equation}
a=\frac{b_1(K-1)}{(K-1)+exp(b_1\tau)}
\end{equation}
Therefore, the value of $b_2=-1$ initiates usual logistic growth whereas $b_2=0.25$ indicates West-type biological growth in the proposed phenomenological description. The variation of specific growth rate with time in case of logistic growth as well as environment dependent West-type biological growth has been shown graphically in figures $7-10$. In case of logistic growth, it is found that time rate of change of specific growth rate changes with $b_1$ and $K$. Rate of this change increases with the increase of $b_1$ and $K$. The higher value of $b_1$ initiates higher initial specific growth rate, but it significantly lowers the time required to attain saturation level, as shown in figure $7$. The variation of $K$ on time required to attain saturation level is not much significant (shown in figure $8$) as it is found in the case of $b_1$. Similar nature is found in case of variation of $b_1$ for the environment dependent West-type biological growth (shown in figure $9$). It is found in figure $10$ that the effect of change of $K$ on specific growth rate is negligible in this case. The specific growth rate corresponding to the environment independent West-type biological growth equation with $\varphi=b_1+b_2a^2$ and $y(0)=a(0)=1$ can be expressed as,
\begin{equation}
a=\frac{b_1}{(b_1+0.25)exp(b_1\tau)+0.25}
\end{equation}
When $b_1=1$, it leads to the environment independent West-type biological growth proposed by Castorina $et$ $al.$ {\cite {Castorina}}. But the other values of $b_1$ ($b_1>0$) also leads to environment independent West-type biological growth equation. The rate of change of specific growth rate increases with the increase of $b_1$ in this case, as shown in figure $11$.\\
Therefore, it can be concluded from phenomenological point of view that two distinct classes of biological growth, following West-type growth equation, are possible to be observed in nature. One of them, proposed by Castorina $et.$ $al.$, is capable of attaining the carrying capacity and is independent of environmental conditions. It is totally governed by the phenomenological coefficients responsible for different growth mechanisms {\cite{Castorina}}. The other, proposed in this communication, does not attain the carrying capacity and is controlled by environmental conditions. Such a classification which is based on the initial growth rate, may or may not be governed by the environmental conditions. One of them is characterized by the condition $\varphi=b_1a+b_2a^2$ with $y(0)=a(0)=1$ (a special case for $b_1=1$ is proposed by Castorina $et$ $al.$ {\cite {Castorina}}). In this type of growth, the functional dependence of specific growth rate ($a$) on $y$ is as follows,
\begin{equation}
a=\frac{b_1(1-y^{b_2})+b_2}{b_1+(b_2-b_1)y^{b_2}}
\end{equation}
Different values of $b_1$ form several subclasses in this category. Another one is following the conditions $\varphi=b_1a+b_2a^2$ with $y(0)=1$ and $a(0)=b_1(1-\frac{1}{K})$, as proposed in this communication.
\begin{figure}
  \centering
  \includegraphics[width=3in,height=2in]{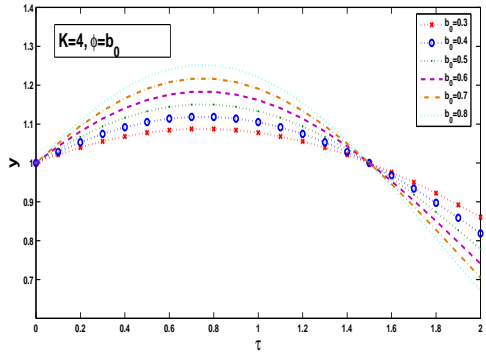}\\
  \caption{(Colour online) Growth curves corresponding to $\varphi=b_0$ with $K=4$. From top to the bottom the values of the parameter $b_0$ are $0.8$, $0.7$, $0.6$, $0.5$, $0.4$ and $0.3$.}
\end{figure}
\begin{figure}
  \centering
  \includegraphics[width=3in,height=2in]{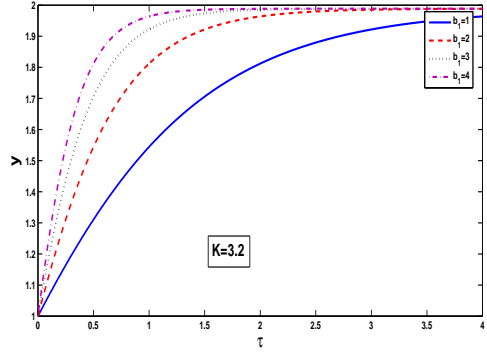}\\
  \caption{(Colour online) Growth curves corresponding to $\varphi=b_1a$ with $K=3.2$. From top to the bottom the values of the parameter $b_1$ are $4$, $3$, $2$ and $1$.}
\end{figure}
\begin{figure}
  \centering
  \includegraphics[width=3in,height=2in]{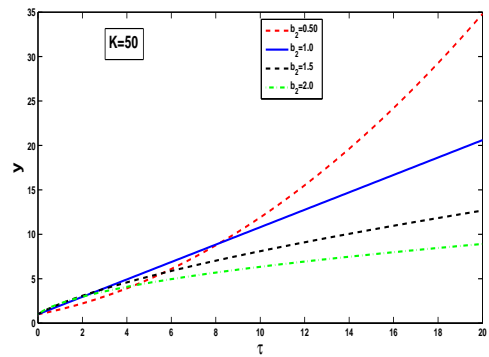}\\
  \caption{(Colour online) Growth curves corresponding to $\varphi=b_2a^2$ with $K=50$. From top to the bottom the values of the parameter $b_2$ are $0.5$, $1.0$, $1.5$ and $2.0$. The solid line represents a straight line.}
\end{figure}
\begin{figure}
  \centering
  \includegraphics[width=3in,height=2in]{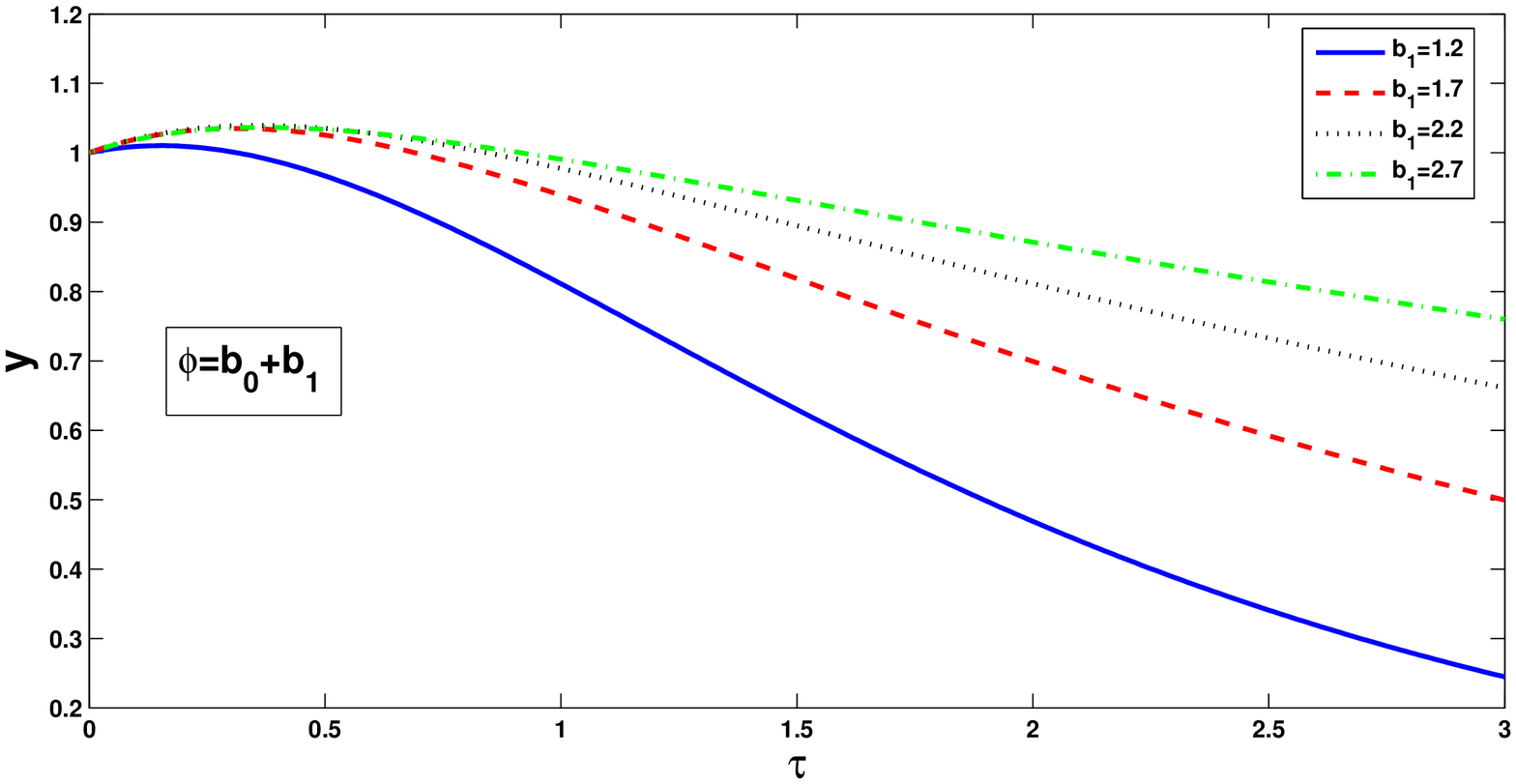}\\
  \caption{(Colour online) Growth curves corresponding to $\varphi=b_0+b_1a$ with $b_0b_1=1$. From top to the bottom the values of the parameter $b_1$ are $2.7$, $2.2$, $1.7$, and $1.2$.}
\end{figure}
\begin{figure}
  \centering
  \includegraphics[width=3in,height=2in]{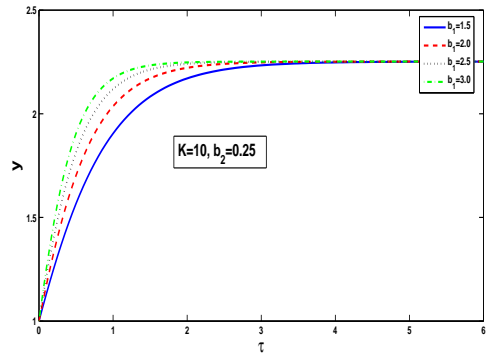}\\
  \caption{(Colour online) Growth curves corresponding to $\varphi=b_1a+b_2a^2$ with $K=5$ and $b_2=0.25$ represent West-type biological growth. From top to the bottom the values of the parameter $b_1$ are $3.0$, $2.5$, $2.0$, and $1.5$. }
\end{figure}
\begin{figure}
  \centering
  \includegraphics[width=3in,height=2in]{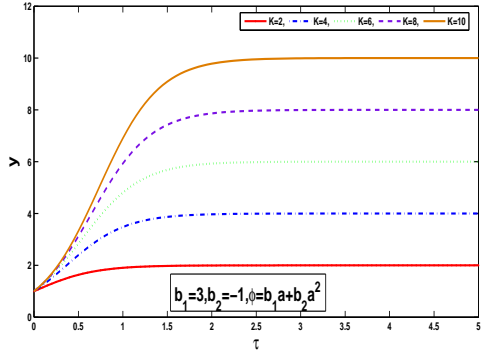}\\
  \caption{(Colour online) Growth curves corresponding to $\varphi=b_1a+b_2a^2$ with $b_1=3$ and $b_2=-1$ (logistic growth). From top to the bottom the values of the parameter $K$ are $10$, $8$, $6$, $4$ and $2$.}
\end{figure}

\begin{figure}
  \centering
  \includegraphics[width=3in,height=2in]{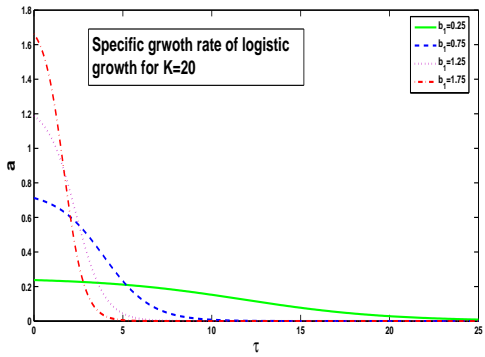}\\
  \caption{(Colour online) Curves of specific growth rate corresponding to logistic growth with $b_2=-1$ and $K=20$. From top to the bottom the values of the parameter $b_1$ are $1.75$, $1.25$, $0.75$, and $0.25$.}
\end{figure}
\begin{figure}
  \centering
  \includegraphics[width=3in,height=2in]{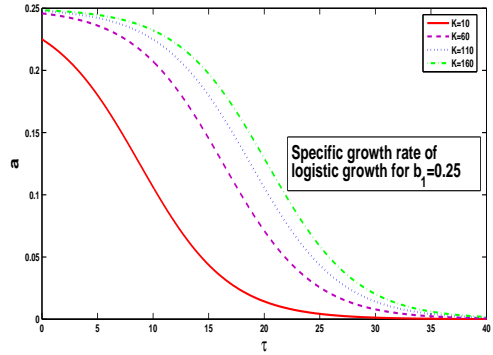}\\
  \caption{(Colour online) Curves of specific growth rate corresponding to logistic growth with $b_1=0.25$ and $b_2=-1$. From top to the bottom the values of the parameter $K$ are $160$, $110$, $60$, and $10$.}
\end{figure}
\begin{figure}
  \centering
  \includegraphics[width=3in,height=2in]{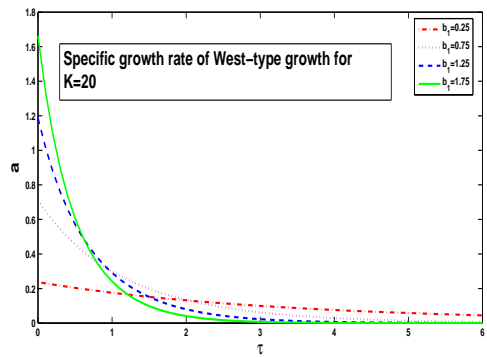}\\
  \caption{(Colour online) Curves of specific growth rate (corresponding to West-type biological growth equation) with $b_2=0.25$ and $K=20$. From top to the bottom the values of the parameter $b_1$ are $1.75$, $1.25$, $0.75$, and $0.25$.}
\end{figure}
\begin{figure}
  \centering
  \includegraphics[width=3in,height=2in]{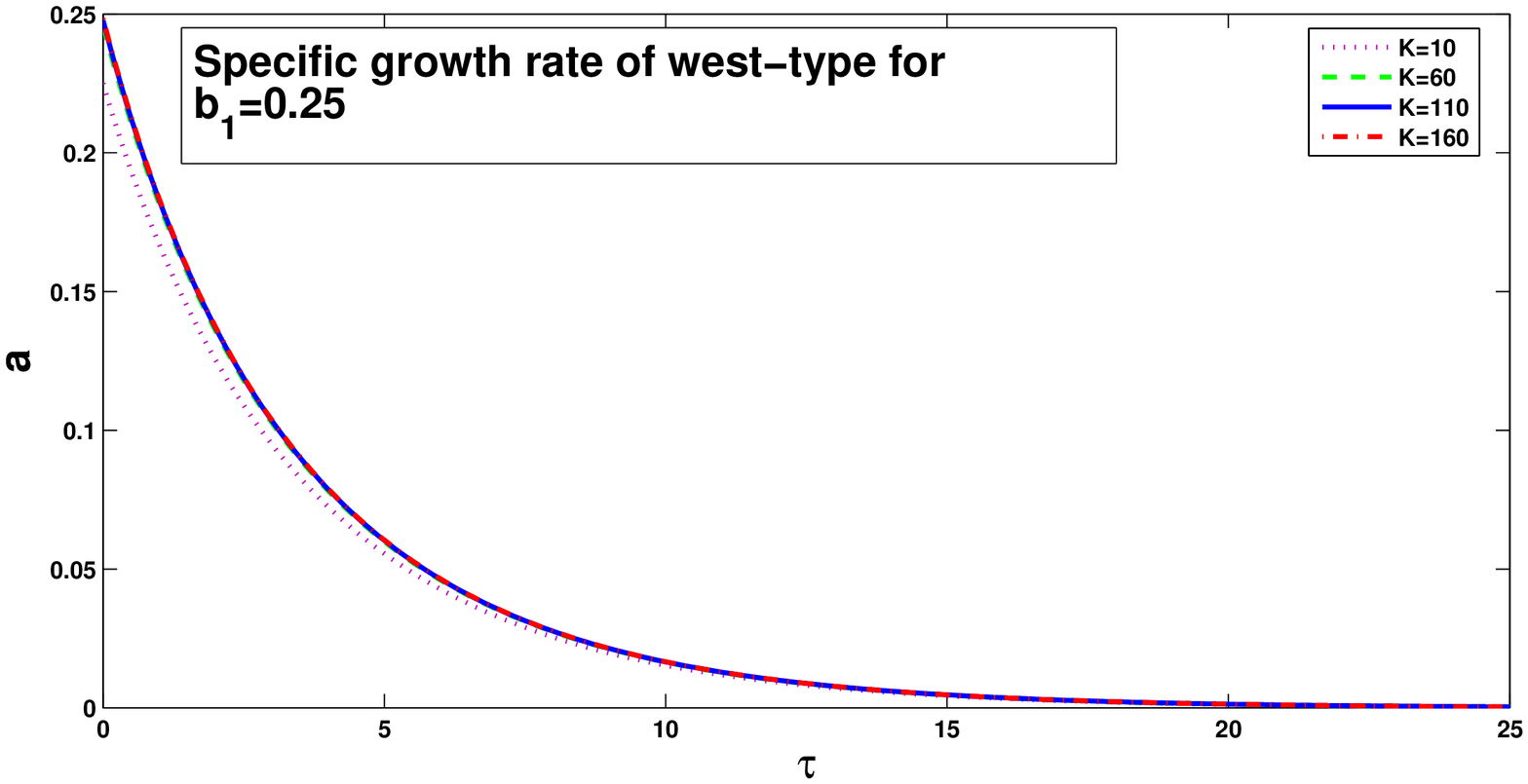}\\
  \caption{(Colour online) Curves of specific growth rate (corresponding to West-type biological growth equation) with $b_1=0.25$, $b_2=0.25$ and $K=20$. From top to the bottom the values of the parameter $K$ are $160$ are $110$, $60$, and $10$.}
\end{figure}
\begin{figure}
  \centering
  \includegraphics[width=3in,height=2in]{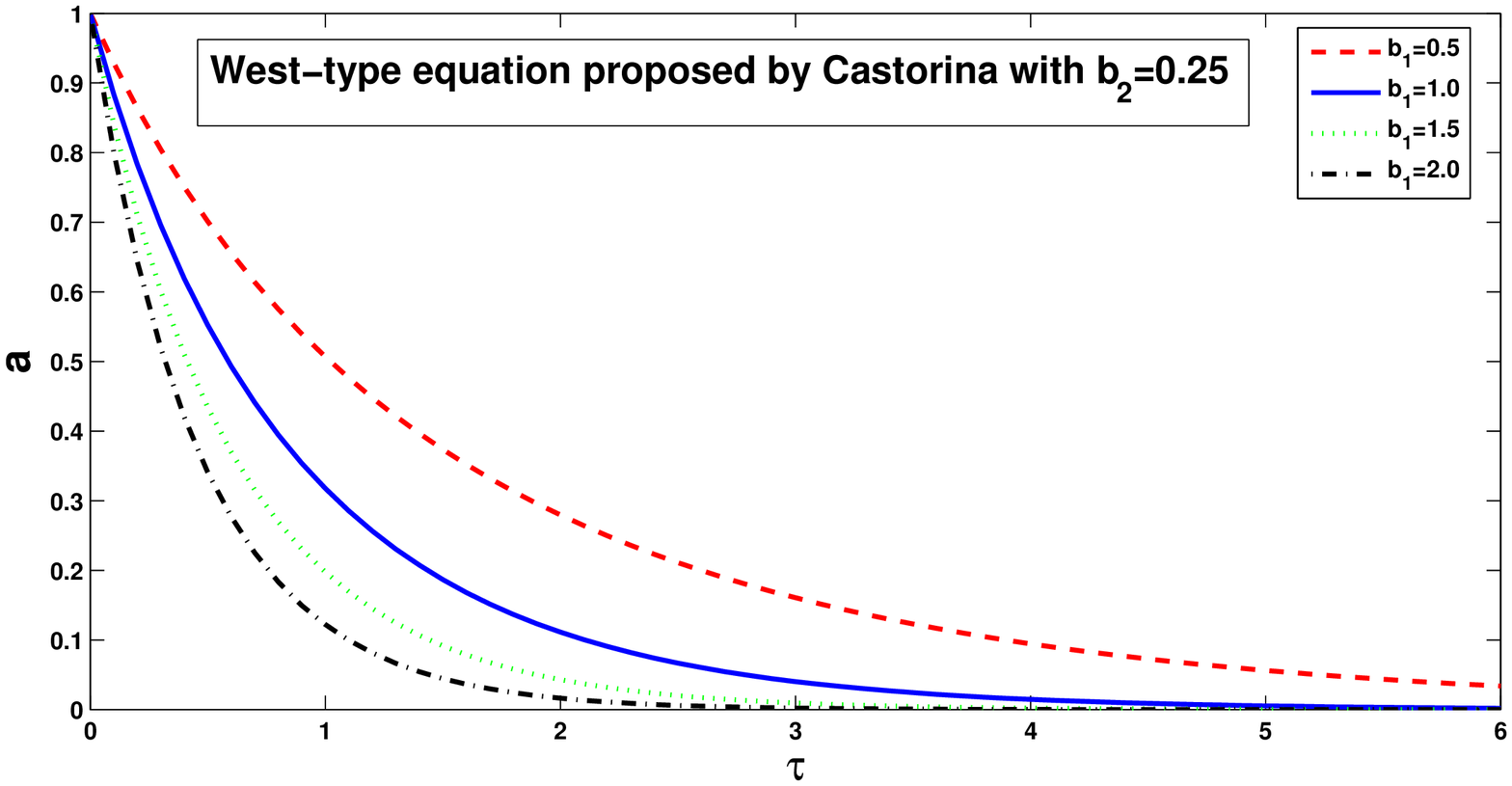}\\
  \caption{(Colour online) Curves of specific growth rate (corresponding to West-type biological growth equation) with $b_2=0.25$. From top to the bottom the values of the parameter $b_1$ are $0.5$, $1.0$, $1.5$, and $2.0$. The solid curve ($b_1=1.0$) corresponds to West-type biological growth proposed by Castorina $et$ $al.$ {\cite {Castorina}}. }
\end{figure}
\section{Conclusions}
Different types of systems are considered in terms of new initial conditions of phenomenological description. Truly linear behaviour has been identified in this study from the phenomenological point of view. It is found that involuted Gompertz function, used to describe growth of a biological system undergoing atrophy or a demographic and economic system undergoing involution or regression, can be addressed in the proposed phenomenological description. The logistic growth and the environment dependent West-type growth of a biological system have been represented in this proposed framework of phenomenological universalities of growth. Another interesting feature is that there is no difference in between these two types of growth in terms of phenomenological approach. The only difference is related to the value of coefficient in phenomenological description. It is also found that two different categories (in a broader sense) of biological growth $-$ one of them is environment independent and the other is environment dependent $-$ lead to the same West-type growth equation. Dependence of growth features on different parameters in each cases is shown graphically. The key observation is that logistic growth and environment dependent West-type growth are originated from the same phenomenological description with different values of phenomenological coefficient. The findings may help the researchers to take a deep insight in the mechanism involved in the biological growth processes.\\

\end{document}